\documentclass[lettersize,journal]{IEEEtran}
\usepackage{amsmath,amssymb,amsfonts}
\usepackage{algorithmic}
\usepackage{algorithm}
\usepackage{array}
\usepackage[caption=false,font=footnotesize,labelfont=rm,textfont=rm]{subfig}
\usepackage{textcomp}
\usepackage{stfloats}
\usepackage{multirow}
\usepackage{makecell}
\usepackage{url}
\usepackage{verbatim}
\usepackage{graphicx}
\usepackage{cite}
\usepackage{hyperref}
\usepackage{booktabs}
\usepackage{wrapfig}
\usepackage{threeparttable}
\hypersetup{hypertex=true,
            colorlinks=true,
            linkcolor=blue,
            anchorcolor=blue,
            citecolor=blue}
\begin{document}

\title{RISAR: RIS-assisted Human Activity Recognition with Commercial Wi-Fi Devices}

\author{Junshuo Liu,~\IEEEmembership{Student Member,~IEEE,}
Tiebin Mi,~\IEEEmembership{Member,~IEEE,}
Xin Shi,~\IEEEmembership{Member,~IEEE,}
Yunlong Huang, Zhe Li, Wei Yang, Rujing Xiong,~\IEEEmembership{Student Member,~IEEE,}
Robert C. Qiu,~\IEEEmembership{Fellow,~IEEE}

\thanks{Manuscript received April xx, 202x; revised August xx, 202x. This work was supported in part by the National Natural Science Foundation of China under Grant No.12141107, and in part by the Interdisciplinary Research Program of HUST, 2023JCYJ012. (Corresponding author: Xin Shi.)}

\thanks{J. Liu, T. Mi, Y. Huang, Z. Li, W. Yang, R. Xiong, and R. Qiu are with the School of Electronic Information and Communications, Huazhong University of Science and Technology, Wuhan 430074, China (e-mail: junshuo\_liu@hust.edu.cn; mitiebin@hust.edu.cn; huangyunlong@hust.edu.cn; m202272435@hust.edu.cn; yangwei\_eic@hust.edu.cn; rujing@hust.edu.cn; caiming@hust.edu.cn).}

\thanks{X. Shi is with the School of Control and Computer Engineering, North China Electric Power University, Beijing 102206, China (e-mail: xinshi@ncepu.edu.cn).}
}

% The paper headers
\markboth{Journal of \LaTeX\ Class Files,~Vol.~14, No.~8, August~2021}%
{Shell \MakeLowercase{\textit{et al.}}: A Sample Article Using IEEEtran.cls for IEEE Journals}

\maketitle

\begin{abstract}
Human activity recognition (HAR) holds significant importance in smart homes, security, and healthcare. Existing systems face limitations because of the insufficient spatial diversity provided by a limited number of antennas. Furthermore, inefficiencies in noise reduction and feature extraction from sensing data pose challenges to recognition performance. This study presents a reconfigurable intelligent surface (RIS)-assisted passive human activity recognition (RISAR) method, compatible with commercial Wi-Fi devices. RISAR leverages a RIS to enhance the spatial diversity of Wi-Fi signals, effectively capturing a wider range of information distributed across the spatial domain. A novel high-dimensional factor model based on random matrix theory is proposed to address noise reduction and feature extraction in the temporal domain. A dual-stream spatial-temporal attention network model is developed to assign variable weights to different characteristics and sequences, mimicking human cognitive processes in prioritizing essential information. Experimental analysis shows that RISAR significantly outperforms existing HAR methods in accuracy and efficiency, achieving an average accuracy of 97.26\%. These findings underscore RISAR's adaptability and potential as a robust activity recognition solution in real environments.
\end{abstract}


\section{Introduction}
\IEEEPARstart{H}uman activity recognition (HAR) is a crucial component in the Internet of Things (IoT) realm. HAR involves the classification of human activities, enabling applications in eldercare, healthcare, and home security \cite{hassan2018robust,zhou2020deep,zhao2022human,lara2012survey}. Recent advancements in Wi-Fi signal-based HAR techniques have highlighted their promising potential. These methodologies often leverage channel state information (CSI) to distinguish various human activities, as demonstrated in systems such as WiSee \cite{pu2013whole}, RT-Fall \cite{wang2016rt}, CARM \cite{wang2017device}, WiAct \cite{yan2019wiact}, and CeHAR \cite{lu2022cehar}. Notably, RT-Fall \cite{wang2016rt} utilized both amplitude and phase data from CSI measurements to identify falls, while CARM \cite{wang2017device} applied CSI-speed and CSI-activity models for recognizing activities. Furthermore, WiAct \cite{yan2019wiact} explored the correlation between body movement and CSI amplitude to classify activities. Nonetheless, these systems face significant challenges, particularly the limited spatial diversity provided by a small number of antennas.

In typical indoor environments, where multiple rooms are connected to a single Wi-Fi access point (AP), Wi-Fi signals are frequently confronted with low signal-noise ratio (SNR) conditions, alongside interference and multipath effects \cite{honma2018human}. Traditional denoising and feature extraction methods, such as principal component analysis (PCA), may not sufficiently address the complexities inherent in Wi-Fi signal data \cite{van2009dimensionality}. Our empirical research indicates that PCA exhibits suboptimal performance in activity recognition. This observation underscores the necessity of exploring more sophisticated techniques to improve HAR performance in these environments.

\begin{figure}[tbp]
\centering
{
\includegraphics[width=1.0\columnwidth]{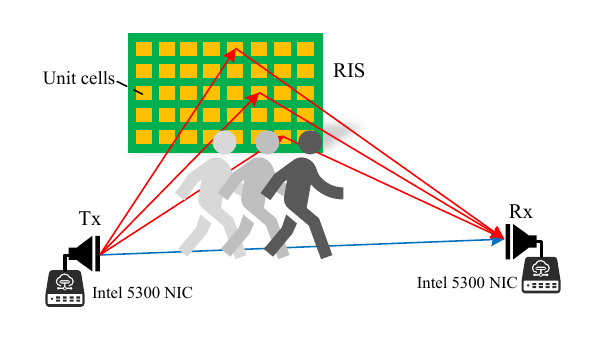}
}
\caption{The schematic of the RIS-based human activity recognition system.}
\label{F1}
\end{figure}

In this study, we introduce a novel reconfigurable intelligent surface-assisted human activity recognition (RISAR) method, which is compatible with commercial Wi-Fi devices. RISAR leverages a RIS to capture a wider range of spatially distributed information, thereby enhancing the spatial diversity of Wi-Fi signals \cite{lan2020wireless,liu2023integrated,rihan2023passive}, as shown in Fig.~\ref{F1}. During the preprocessing phase, we introduce a high-dimensional factor model for extracting activity features from raw CSI measurements in the temporal domain. Additionally, we present a specialized dual-stream spatial-temporal attention network (DS-STAN) architecture tailored to address the nuanced characteristics inherent in sensing data. This architecture is trained to interpret the patterns embedded within CSI data.

The key contributions of our research are delineated as follows:
\begin{itemize}
\item An innovative RISAR approach is presented for human activity recognition under low SNR conditions. The integration of RIS not only introduces additional links but also enhances spatial diversity, leading to a significant improvement in signal quality and system performance. To the best of our knowledge, this represents the first attempt to leverage RIS technology for human activity recognition using commodity Wi-Fi devices.

\item A high-dimensional factor model approach is proposed for extracting activity features from CSI measurements in the temporal domain. These extracted features are fed into the proposed DS-STAN model, which integrates an attention mechanism to discriminate various human activities. This methodology represents a sophisticated fusion of random matrix theory and deep learning techniques, offering enhanced accuracy in activity recognition.

\item The RISAR system, operational with commercial Wi-Fi devices under various conditions, has undergone extensive evaluation. The empirical results highlight its remarkable accuracy, achieving an average of 97.26\% in the Office dataset and 90.83\% in the L-shaped hallway dataset. These tests highlight the robustness and adaptability of RISAR across a range of environmental settings.
\end{itemize}

The remainder of this paper is organized as follows. We start by providing a comprehensive literature review of existing work in Section II. In Section III, we present the system architecture, as well as the preprocessing and recognition algorithms. Section IV illustrates the performance of the proposed approach. Finally, Section V concludes the paper.

\section{Related Work}
In this section, we provide a summary of the relevant literature concerning human activity recognition. There exist three modalities in HAR: vision-based \cite{xu2013exploring,kim2019vision,franco2020multimodal,sharma2022review}, wearable device-based \cite{bhat2018online,hassan2018robust,bianchi2019iot,zhang2022deep}, and wireless sensing-based systems \cite{zhu2022continuous,oguntala2019smartwall,pu2013whole,wang2016rt,wang2017device,yousefi2017survey,chen2018wifi,yan2019wiact,feng2019wi,xiao2020deepseg,guo2021towards,lu2022cehar,shalaby2022utilizing}.

Vision-based methodologies leverage high-resolution cameras to capture images or videos for activity recognition \cite{xu2013exploring}. For instance, Kim \textit{et al.} \cite{kim2019vision} introduced a depth video-based HAR system, utilizing skeletal joint features to discern the daily activities of the elderly in indoor settings. Additionally, Franco \textit{et al.} \cite{franco2020multimodal} amalgamated skeleton and RGB data stream analyses to extract features for HAR. These systems are are sensitive to lighting conditions and obstructions, in addition to security and privacy concerns \cite{sun2022human}.

Wearable device-based HAR approaches utilize technologies such as smartphones \cite{hassan2018robust}, IoT devices \cite{bhat2018online}, and sensors \cite{bianchi2019iot} for activity detection. Hassan \textit{et al.} \cite{hassan2018robust} developed a robust HAR system based on smartphone sensor data, capable of distinguishing transitional and non-transitional activities. Bhat \textit{et al.} \cite{bhat2018online} employed a low-power IoT device for HAR, pioneering a framework for both online training and inference. Bianchi \textit{et al.} \cite{bianchi2019iot} explored the integration of inertial measurement units (IMUs) with deep learning, where wearable sensors transmit data to a cloud service for activity monitoring. These methods suffer from the drawback of requiring long-term wearing, leading to inconvenience among users \cite{sun2022human}.

Wireless sensing-based HAR encompasses technologies like radar, radio frequency identification (RFID), and Wi-Fi signals. Radars, capable of detecting human motion without necessitating user-carried devices, offer resilience against variations in lighting conditions due to their broad bandwidth \cite{islam2022human}. Zhu \textit{et al.} \cite{zhu2022continuous} proposed a distributed ultra-wideband radar-based system, classifying motions through spatio-temporal patterns of radar Doppler signatures. However, limitations in these systems include portability and hardware costs. RFID systems present a cost-effective alternative, requiring minimal hardware such as chip tags and antennas. Oguntala \textit{et al.} \cite{oguntala2019smartwall} introduced SmartWall, a system employing passive RFID tags for the recognition of sequential activities. The RFID-based approach offers a cost-effective alternative but encounters robustness issues in complex indoor environments, primarily due to the multipath effect and channel variations \cite{yang2013rssi}.

In the evolving landscape of Wi-Fi signal-based technologies, a diverse array of systems have been developed for HAR. Yet, these systems frequently necessitate specialized devices and are often inadequate in non-line-of-sight (NLOS) scenarios. For instance, the WiSee system \cite{pu2013whole} confirmed the efficacy of Wi-Fi signals in detecting human motion in NLOS and through-the-wall environments, albeit relying on USRP-N210 devices rather than standard commercial Wi-Fi hardware. Yousefi \textit{et al.} \cite{yousefi2017survey} employed long short-term memory (LSTM) networks to encode temporal information during feature learning, setting a new benchmark in HAR using CSI measurements. Similarly, Chen \textit{et al.} \cite{chen2018wifi} innovatively utilized raw CSI signals in conjunction with an attention-based bi-directional LSTM to ascertain action classes. Feng \textit{et al.} \cite{feng2019wi} proposed a three-phase system for multiple human activity recognition, focusing on identifying the start and end points of activities in multi-subject scenarios. Xiao \textit{et al.} \cite{xiao2020deepseg} introduced a deep learning-based HAR approach that capitalizes on Wi-Fi signals, transforming segmentation tasks into classification challenges via a convolutional neural network (CNN) algorithm. Guo \textit{et al.} \cite{guo2021towards} developed an encoder-decoder framework aimed at mitigating accuracy disparities among individuals in HAR applications. Lu \textit{et al.} \cite{lu2022cehar} in their CeHAR system, advocated for the amalgamation of CSI amplitude and phase features to enhance action detection efficiency. Lastly, Shalaby \textit{et al.} \cite{shalaby2022utilizing} presented a deep learning model integrating CNNs, gated recurrent units (GRUs), and LSTMs, designed to extract high-dimensional and time sequence features.

These systems generally do not consider HAR in NLOS environments. This is particularly pertinent in many indoor settings, where multiple rooms often connect to a single wireless signal AP, especially in residential environments. In such scenarios, issues like signal attenuation and multipath fading can significantly impair the accuracy of activity recognition.

\section{System Architecture}
\begin{figure*}[tbp]
  \centering
  \includegraphics[width=1.9\columnwidth]{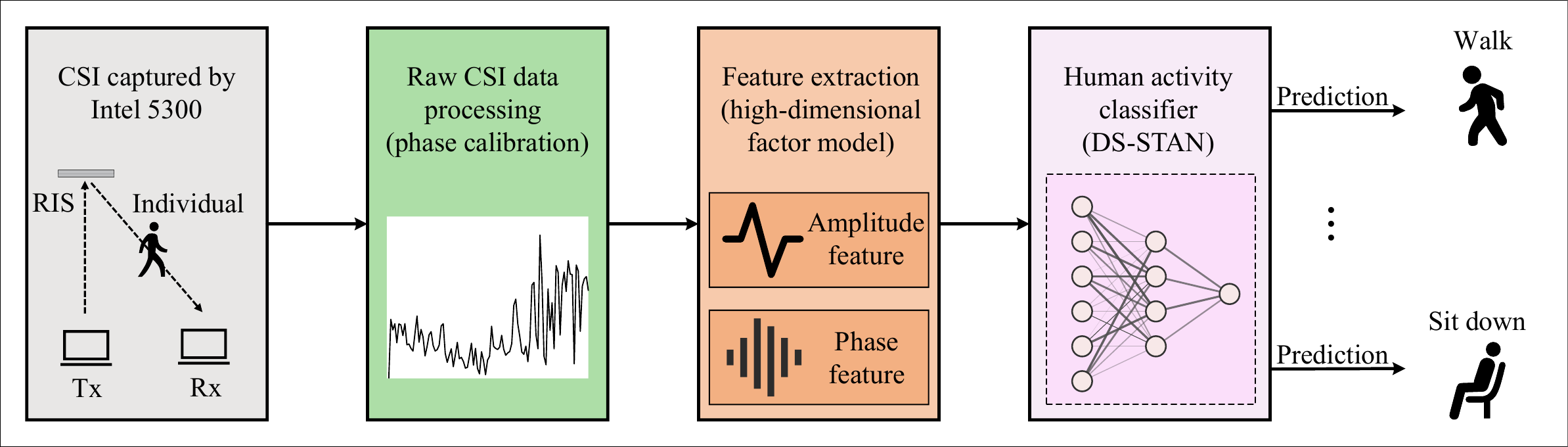}
  \caption{The architecture of the proposed RISAR system.}
  \label{F2}
\end{figure*}

This section focuses on the details of the RISAR. The integration of RIS for HAR not only introduces additional links but also enhances spatial diversity. The architecture of the RISAR system is illustrated in Fig.~\ref{F2}. Following the data flow, the data processing procedure consists of three stages: CSI data collection, data preprocessing, and the human activity classification algorithm.

\subsection{Hardware Architecture and Testing Settings}
A 1-bit RIS utilized in this study operates at a frequency of 5.8 GHz. It consists of $16 \times 10$ reflecting units, with each unit spaced $0.4 \lambda$ apart. To simplify the experiments, all devices are positioned at the same height. We utilize two Intel Wi-Fi Link 5300 network interface cards (NICs), serving as both the transmitter (Tx) and receiver (Rx).

\begin{figure*}[htbp]
\centering
\subfloat[L-shaped hallway.]{
\label{F3-a}
\includegraphics[width=0.82\columnwidth]{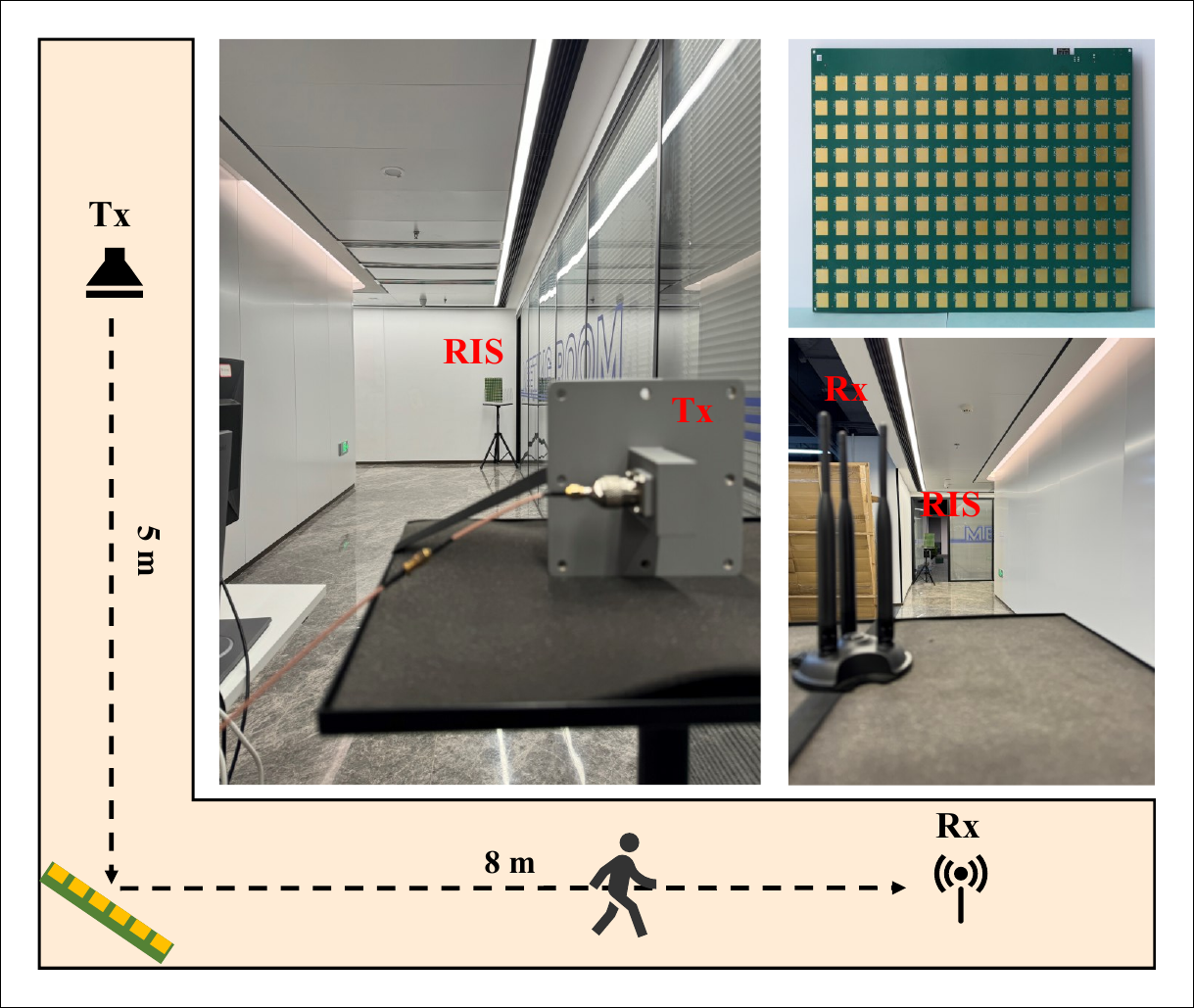}
}
\subfloat[Floor plans of the other two experimental environments.]{
\label{F3-b}
\includegraphics[width=1.1\columnwidth]{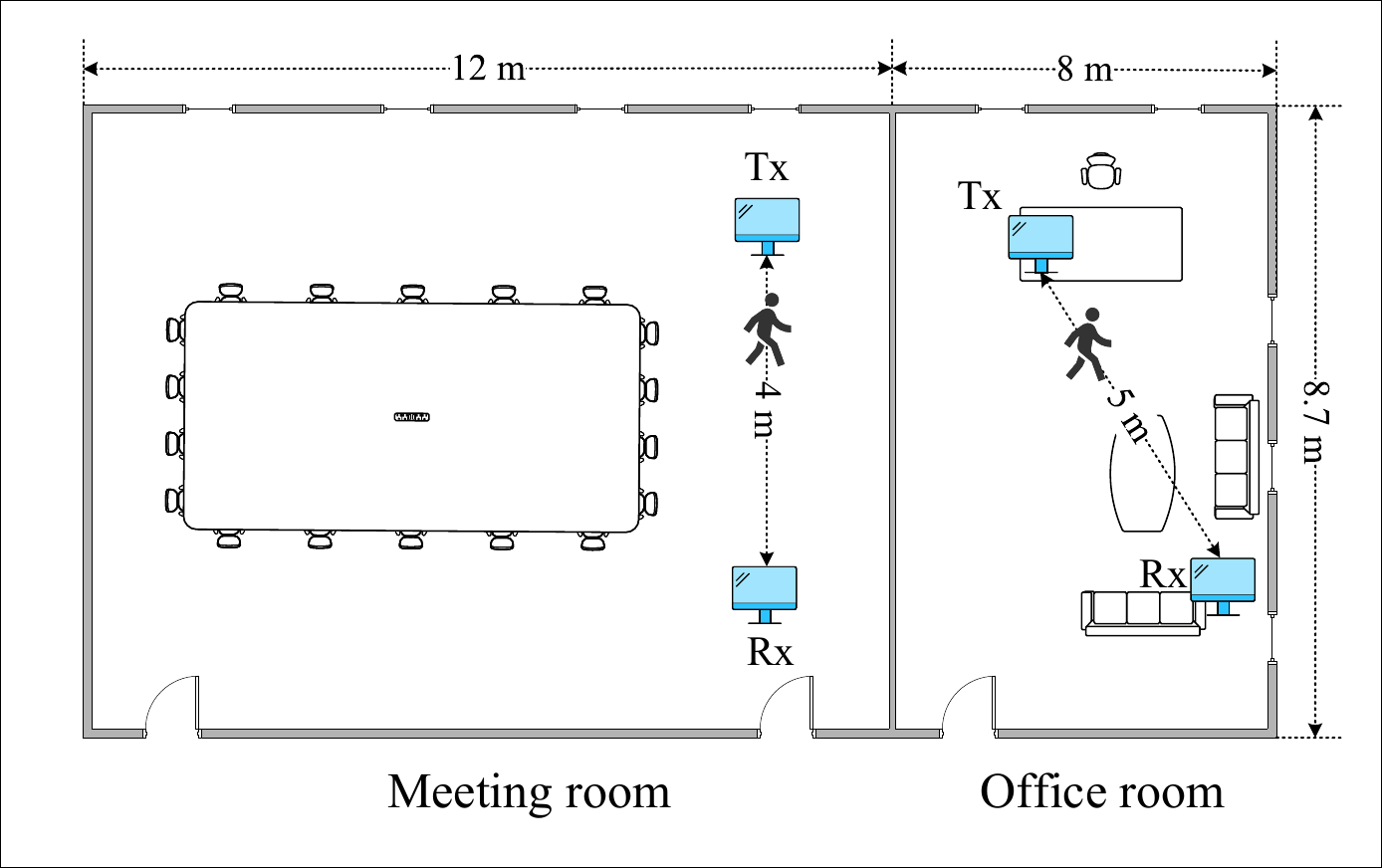}
}
\caption{Three types of CSI data collection environments.}
\label{F3}
\end{figure*}

To evaluate the effectiveness of RISAR in an extremely low SNR environment, initial tests are conducted within an L-shaped hallway, as depicted in Fig.~\ref{F3}~\subref{F3-a}. The setup involves an Intel 5300 NIC with a horn antenna serving as the Tx, while an Intel 5300 NIC equipped with three omnidirectional antennas acts as the Rx. Positioned at the corner, the RIS functions as a relay to enhance the diversity of the environment. Utilizing the DaS algorithm \cite{xiong2022optimal}, the RIS is optimized to achieve optimal phase configurations. Its efficiency is validated in our prior study \cite{xiong2023ris}, yielding an improvement of at least 10 dBi.

Moreover, to validate the role of the proposed data preprocessing and recognition algorithms, two additional tests are conducted in a meeting room and an office room. Unlike the previous testing scenario in the L-shaped hallway, RISs are not utilized in these setups. This configuration is implemented to isolate the impact of the proposed algorithms. The tests, as illustrated in Fig.~\ref{F3}~\subref{F3-b}, involve data collection in a meeting room covering a $12 \times 8.7 \, \text{m}^2$ area and an office room spanning an $8 \times 8.7 \, \text{m}^2$ space. Both the Tx and Rx are configured with three antennas. The Intel 5300 NIC is set to collect data on the 161-st channel with a 20 MHz channel bandwidth.

\subsection{Data Collection}
To capture the instantaneous CSI, which exhibits sensitivity to human activities, this study engaged in the analysis of packets harvested by the Intel Wi-Fi Link 5300 NIC within the temporal domain. Each packet is composed of a matrix characterized by dimensions $N_{\text{Tx}} \times N_{\text{Rx}} \times N_{\text{sc}}$, where $N_{\text{Tx}}$ and $N_{\text{Rx}}$ signify the count of transmitting and receiving antennas, respectively, and $N_{\text{sc}}$ indicates the orthogonal frequency-division multiplexing (OFDM) subcarriers \cite{halperin2011tool}. Specifically, the Intel Wi-Fi Link 5300 NIC outputs CSI data for 30 subcarriers corresponding to each of its three antennas. Consequently, during the tests in the L-shaped hallway, each packet generates a data frame containing $1 \times 3 \times 30$ CSI measurements. By recording 5000 packets, we obtain a data frame of size $90 \times 5000$, representing one activity in the dataset. On the other hand, in the meeting room and office room tests, the data frame consists of $3 \times 3 \times 30$ CSI measurements. Therefore, each activity is of size $270 \times 5000$.

RISAR focuses on recognizing human daily activities. We select 10 typical activities for examination, as delineated in Table~\ref{table1}. Participants repeated each activity 100 times to build a comprehensive dataset. Each record is sampled at a rate of 1000 packets per second over a 5-second interval.

\begin{table}[htbp]
\centering
\caption{Typical categories of activities tested in the experiments.}
\setlength{\tabcolsep}{3.5mm}
\begin{threeparttable}
\begin{tabular}{cccc}
\toprule
\multirow{2}{*}{Human activity} & \multicolumn{3}{c}{Number of samples} \\ \cmidrule(lr){2-4}
& L-shaped hallway  & Meeting room & Office    \\
\midrule
bend          & 400 & -   & 600        \\
empty         & 400 & -   & -          \\
fall          & 400 & -   & 600        \\
lie down      & -   & 600 & -          \\
pick up       & -   & 600 & -          \\
run           & -   & -   & 600        \\
sit down      & 400 & 600 & 600        \\
stand         & -   & 600 & 600        \\
stand up      & 400 & 600 & -          \\
walk          & -   & 600 & 600        \\
wave          & 400 & -   & 600        \\
\bottomrule
\end{tabular}
\begin{tablenotes}
\item - denotes there is no sample in the dataset.
\end{tablenotes}
\end{threeparttable}
\label{table1}
\end{table}

\subsection{Data Analysis and Preprocessing}
The utilization of low-cost Wi-Fi cards for data collection introduces inherent imperfections in the CSI data, underscoring the necessity for preliminary data cleaning processes \cite{ma2019wifi}. Moreover, the substantial dimensionality inherent to each data frame necessitates the development of data compression algorithms to enhance the efficiency of model training processes. In the context of learning-based recognition algorithms, the process of dimension reduction serves as a critical intermediary feature extraction technique.

\subsubsection{Activity feature analysis}
\begin{figure*}[tbp]
  \centering
  \includegraphics[width=1.7\columnwidth]{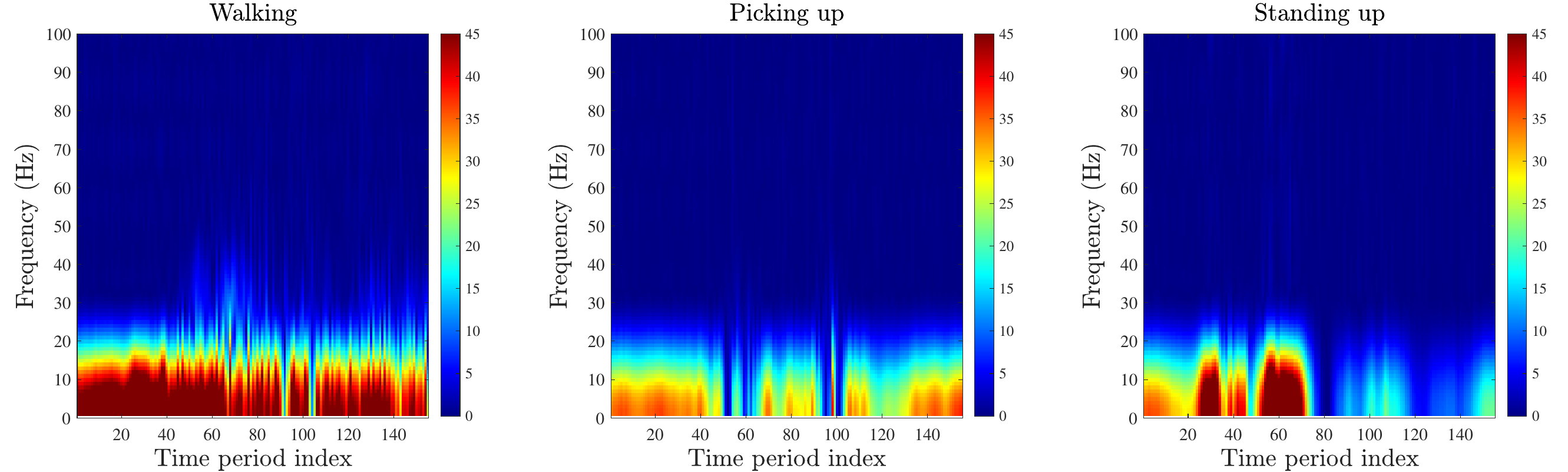}
  \caption{The spectrogram of one subcarrier's CSI amplitude for different activities: walking, picking up, and standing up (from left to right).}
  \label{F4}
\end{figure*}

The principle that CSI encapsulates the cumulative effects of scattering, fading, and power decay within the channels is well-recognized. Consequently, the occurrence of varied human activities induces detectable fluctuations in the CSI, forming the basis for activity recognition methodologies. For enhanced elucidation of this phenomenon, the application of the short-time Fourier transform (STFT) on the CSI data is employed to delineate features across both temporal and frequency domains \cite{wang2016rt,wang2017device}. Fig.~\ref{F4} showcases the STFT (spectrogram) of the CSI corresponding to diverse activities, elucidating that actions characterized by pronounced movements, such as walking, manifest significantly elevated energy levels in the spectrogram. This pivotal time-frequency data proves essential for the accurate recognition of distinct human activities.

\subsubsection{Data cleaning and compression}
To address low-level imperfections such as phase offset in CSI measurements, which include sampling time offset (STO) and sampling frequency offset (SFO), we employ a linear transformation method \cite{qian2014pads}. This method effectively mitigates shifts caused by STO/SFO, resulting in more accurate phase values.

Within the domain of data compression, previous research \cite{wang2015understanding} has validated the effectiveness of principal component analysis. One disadvantage of using PCA for compression lies in its assumption of linearity within the dataset. However, the CSI measurements often contain nonlinear patterns. Furthermore, the CSI data exhibit temporal correlations ($T$ sampling packets) in conjunction with spatial correlations ($N$ subcarriers). PCA may not adequately capture these complex dependencies within the data.

To address the temporal-spatial correlations in the CSI measurements, random matrix theory offers a mathematically rigorous solution. We propose a novel denoising and feature extraction algorithm known as the high-dimensional factor model (HDFM), which is based on the Marčenko-Pastur (M-P) law \cite{marchenko1967distribution}, one of the most remarkable laws in the random matrix theory.

Recall that $N$ represents the number of subcarriers and $T$ is the number of time sampling points. Let $\mathbf{R}_{it}$ denote the amplitude/phase of the $i$-th subcarrier at time $t$. Under the Kolmogorov condition, i.e., $T, N \to \infty$ and $\frac{N}{T} \to c > 0$, the CSI amplitude/phase matrix can be represented by Eq.~\eqref{factormodel}. Here $\mathbf{F}_{jt}$ denotes the $j$-th temporal factor at time $t$, $\mathbf{L}_{ij}$ represents the loading of the $j$-th temporal factor on the $i$-th cross-sectional element, $\mathbf{U}_{it}$ stands for the additive noise, and $p$ is the number of temporal factors.
\begin{equation}
\label{factormodel}
\mathbf{R}_{it}=\sum_{j=1}^{p}\mathbf{L}_{ij}\mathbf{F}_{jt}+\mathbf{U}_{it}.
\end{equation}

We construct the $p$-level residuals by removing temporal factors, using principal components:
\begin{equation}
\label{residual}
\mathbf{U}_{it}^{(p)} = \mathbf{R}_{it} - \mathbf{L}_{ij}^{(p)} \mathbf{F}_{jt}^{(p)},
\end{equation}
where $\mathbf{L}^{(p)} \mathbf{F}^{(p)}$ represents the estimated common factor obtained from $p$ principal components.

In the proposed HDFM method, we examine the eigenvalue distribution of the covariance matrix of $p$-level residuals $\mathbf{U}_{it}^{(p)}$:
\begin{equation}
\label{covariance}
\mathbf{C}_{it}^{(p)} = \frac{1}{T} \mathbf{U}_{it}^{(p)} \mathbf{U}_{it}^{(p) H}.
\end{equation}

The assumption is that if the factors are accurately eliminated, the residual spectral density $\mathbf{C}_{it}^{(p)}$ closely conforms to the Marčenko-Pastur law. This enables us to isolate the main factors present in the CSI data. For instance, by setting $p=8$, we compress the original CSI data frame from $270 \times 5000$ to $8 \times 5000$. The pseudocode for the proposed method is shown in Algorithm~\ref{Algo1}. For an in-depth proof of the HDFM, please refer to Appendix~\ref{proof}.

\begin{algorithm}
 \caption{The HDFM method.}
 \label{Algo1}
 \begin{algorithmic}[1]
 \renewcommand{\algorithmicrequire}{\textbf{Input:}}
 \renewcommand{\algorithmicensure}{\textbf{Output:}}
 \REQUIRE CSI amplitude/phase data matrix: $\mathbf{R}_{it}$;
 \ENSURE  CSI amplitude/phase features (temporal factors) $\mathbf{F}_{jt}^{(p)}$.
  \STATE Initialize the $p$-level residual matrix by Eq.~\eqref{residual};
  \STATE Compute the covariance matrix by Eq.~\eqref{covariance};
  \STATE Calculate the eigenvalues $\mathbf{\lambda}$ of $\mathbf{C}_{it}^{(p)}$;
  \STATE Compute the eigenvalue distribution $\rho_{\mathbf{\lambda}} (p)$;
  \STATE Generate a Gaussian noise matrix $\hat{\mathbf{U}}_{it}\sim \mathcal{N}(0,\sigma^2)$;
  \STATE Compute the covariance matrix for noise by Eq.~\eqref{covariance};
  \STATE Compute the eigenvalues $\hat{\mathbf{\lambda}}$ of $\hat{\mathbf{C}}_{it}$;
  \STATE Compute the eigenvalue distribution $\rho_{\hat{\mathbf{\lambda}}} (\sigma)$;
  \STATE Minimize the distance between the two spectral distributions;
  \STATE Determine the number of temporal factors $p$ ;
  \STATE Compute the temporal factors $\mathbf{F}_{jt}^{(p)}$ as the final feature representation.
 \end{algorithmic} 
\end{algorithm}

\subsection{Learning-based Algorithms}
In this subsection, we integrate both machine learning (ML) and deep learning (DL) methodologies to evaluate the HDFM. We employ well-established ML algorithms like random forest (RF) \cite{xu2017human}, extreme gradient boosting (XGBoost) \cite{zhang2019comprehensive}, and support vector machine (SVM) \cite{shuvo2020hybrid}, alongside a novel dual-stream spatial-temporal attention deep learning framework designed for human activity classification.

\begin{figure*}[tbp]
  \centering
  \includegraphics[width=1.7\columnwidth]{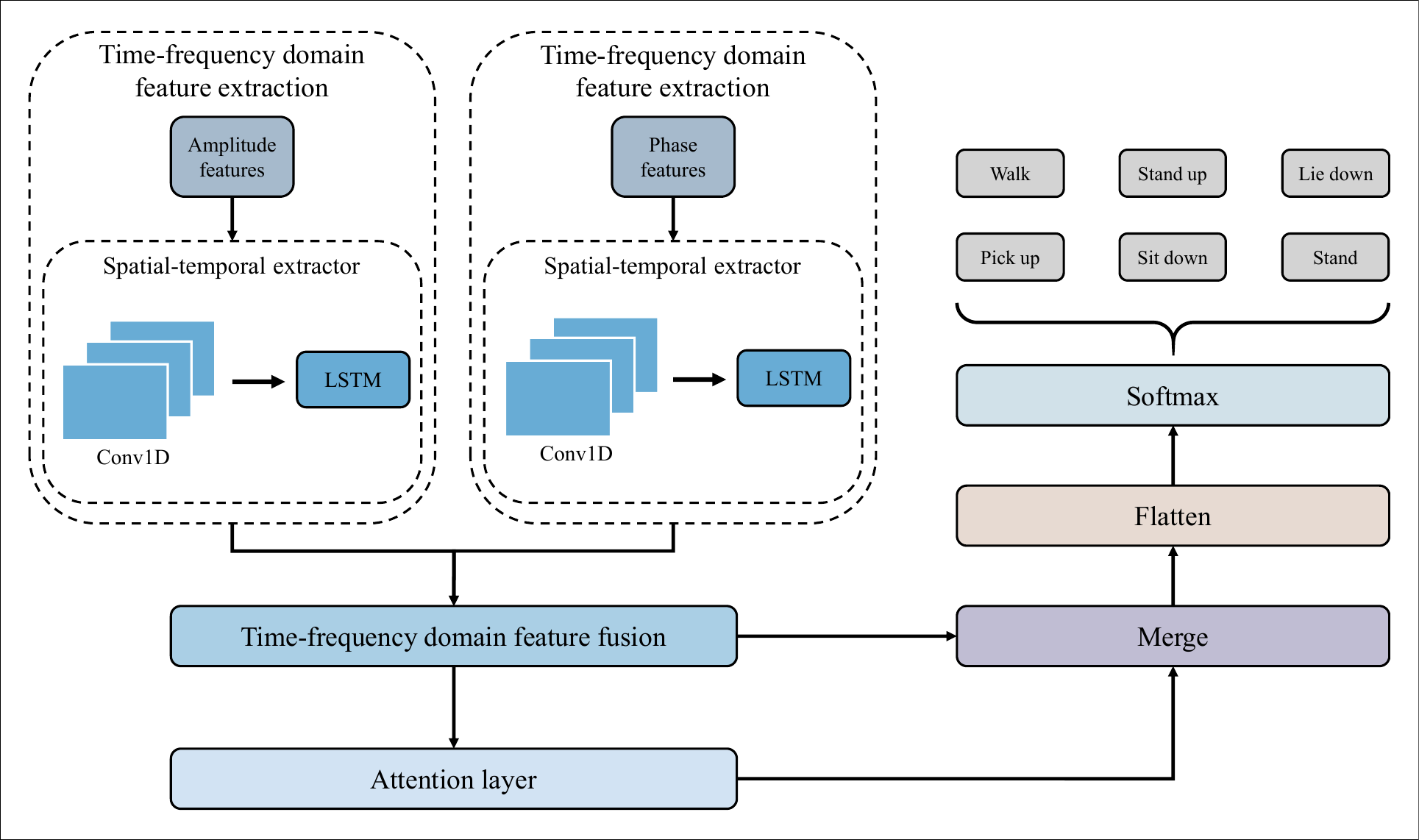}
  \caption{Schematic of our proposed dual-stream spatial-temporal attention deep learning framework. Given CSI amplitude and phase features, we first use the spatial-temporal extractors to extract its activity-related features. Then, these features are fused using the attention mechanism and passed to a dense layer. Finally, the softmax is used to give output for the activity category prediction.}
  \label{F5}
\end{figure*}

This framework's architecture, as depicted in Fig.~\ref{F5}, begins with the preprocessing of CSI data through a spatial-temporal extractor. Following this, the extracted features are input into an attention layer, which utilizes an attention matrix to highlight the significance of various features and time steps. The integration of these features with the attention matrix elements results in a feature matrix with attention weights, which is then flattened into a feature vector. This vector facilitates the differentiation of action labels through a softmax classification layer.

The attention mechanism model, initially developed for computer vision and based on the human ability to selectively concentrate on specific information, is adeptly applied in Wi-Fi-based passive sensing \cite{niu2021review}. The high dimensionality of the features and the varying contributions of different features and time steps to the sensing task are managed through the incorporation of the attention mechanism, as substantiated by our time-frequency analysis illustrated in Fig.~\ref{F4}. The integration of this mechanism significantly enhances the accuracy of Wi-Fi sensing in our study.

In the architecture proposed, the network incorporates three convolutional layers followed by an LSTM layer, which serves to further refine the feature extraction process. As delineated in Table~\ref{table2}, detailed specifications for each layer are provided. Specifically, the convolutional layers are configured with 32, 64, and 128 neurons, respectively, while the LSTM layer is equipped with 512 hidden units. The variable $M$ denotes the batch size, and $K$ signifies the number of labels.

\begin{table}[htbp]
\centering
\caption{Implementation details of the proposed network.}
\setlength{\tabcolsep}{4mm}
\begin{tabular}{cccc}
\toprule
\multirow{2}{*}{Layers} & \multicolumn{2}{c}{Output dimension}  \\
\cmidrule(lr){2-3} & Amplitude features & Phase features   \\
\midrule
Input layer             & ($M$, 8, 5000)    & ($M$, 8, 5000)  \\
Conv1D 1                & ($M$, 32, 5000)   & ($M$, 32, 5000) \\
MaxPool1D 1             & ($M$, 32, 2500)   & ($M$, 32, 2500) \\
Conv1D 2                & ($M$, 64, 2500)   & ($M$, 64, 2500) \\
MaxPool1D 2             & ($M$, 64, 1250)   & ($M$, 64, 1250) \\
Conv1D 3                & ($M$, 128, 1250)  & ($M$, 128, 1250) \\
MaxPool1D 3             & ($M$, 128, 625)   & ($M$, 128, 625)  \\
LSTM                    & ($M$, 625, 512)   & ($M$, 625, 512)   \\
Concat                  & \multicolumn{2}{c}{($M$, 625, 1024)}  \\
Attention layer         & \multicolumn{2}{c}{($M$, 1024)}      \\
Dense layer             & \multicolumn{2}{c}{($M$, $K$)}    \\
\bottomrule
\end{tabular}
\label{table2}
\end{table}

\section{Experiments and Results}
In this section, we implement and evaluate the performance of different methods as well as our proposed dual-stream spatial-temporal attention network model. All experiments are conducted on a server equipped with an Intel Xeon Gold 6138 CPU @ 2.00 GHz, 128 GB RAM, NVIDIA GeForce RTX 3090, operating on Linux and employing Python 3.8 and PyTorch 1.13.

Recognition performance can be measured from different perspectives using the following metrics: 1) Accuracy; 2) Precision; 3) Recall; and 4) F1-score. These metrics are defined as follows:
\begin{equation}
\label{accuracy}
\text{Accuracy}=\frac{\text{TP}+\text{TN}}{\text{TP}+\text{TN}+\text{FP}+\text{FN}},
\end{equation}
\begin{equation}
\label{precision}
\text{Precision}=\frac{\text{TP}}{\text{TP}+\text{FP}},
\end{equation}
\begin{equation}
\label{recall}
\text{Recall}=\frac{\text{TP}}{\text{TP}+\text{FN}},
\end{equation}
\begin{equation}
\label{f1score}
\text{F1-score}=2\times\frac{\text{Precision}\times\text{Recall}}{\text{Precision}+\text{Recall}},
\end{equation}
where $\text{TP}$, $\text{TN}$, $\text{FP}$, and $\text{FN}$ denote true positive, true negative, false positive, and false negative, respectively.

\subsection{Performance Evaluation of RIS-enabled HAR}
In this subsection, a comparative assessment is conducted to evaluate the efficacy of various algorithms within two distinct settings: environments without RIS configuration and those with RIS-optimized configurations, as elaborated in Table~\ref{table3}. This examination highlights substantial enhancements in classifier accuracy following environmental optimization via RIS, with our proposed DS-STAN experiencing a significant accuracy uplift from 79.17\% in non-configurable settings to 90.83\% in RIS-optimized conditions. 

Comparable improvements are observed across other methodologies upon RIS activation, validating the critical role of RIS in augmenting the performance of activity-sensing technologies. Such advancements underscore the potential of RIS in revolutionizing wireless sensing systems, offering a pathway to achieve heightened precision and reliability in outcomes.

\begin{table*}[tbp]
\centering
\caption{The overall recognition performances of different classification methods on non-configurable environment case and optimized-configurable environment case.}
\begin{tabular}{ccccccccc}
\toprule
\multirow{2}{*}{Methods} & \multicolumn{4}{c}{RIS non-configurable environment} & \multicolumn{4}{c}{RIS optimized-configurable environment}         \\
\cmidrule(lr){2-9}
& Accuracy & Recall & Precision & F1-score 
& Accuracy & Recall & Precision & F1-score              \\
\midrule
RF-HDFM       & 0.5083 & 0.5148 & 0.5000 & 0.4837 
              & 0.8583 & 0.8649 & 0.8548 & 0.8547        \\
XGBoost-HDFM  & 0.5292 & 0.5256 & 0.5265 & 0.5214
              & 0.8417 & 0.8417 & 0.8364 & 0.8378        \\
SVM-HDFM      & 0.5208 & 0.5200 & 0.5114 & 0.5116 
              & 0.8583 & 0.8598 & 0.8520 & 0.8546         \\
DS-STAN-HDFM     & 0.7917 & 0.7907 & 0.7881 & 0.7859         
              & \textbf{0.9083} & 0.9084 & 0.9094 & 0.9089  \\
\bottomrule
\end{tabular}
\label{table3}
\end{table*}

\subsection{Performance Evaluation of HDFM}
In the preprocessing phase for CSI, PCA \cite{wang2017device} and HDFM techniques are utilized for noise reduction and feature extraction. The extracted amplitude and phase features are subsequently input into various classifiers, including RF, XGBoost, SVM, and the proposed DS-STAN model that integrates time-frequency domain features. A comprehensive evaluation reveals HDFM's significant enhancement of activity recognition accuracy in two datasets, with noticeable improvements over PCA-based approaches, as detailed in Table~\ref{table4}. Notably, the implementation of HDFM significantly enhances recognition accuracy in the Meeting room dataset, with improvements observed between 5.14\% and 5.70\% over PCA methods. Similarly, for the Office dataset, HDFM outperforms PCA, indicating accuracy gains ranging from 6.83\% to 9.48\%. It validates that the features calculated through our proposed HDFM can be more representative of the CSI data than those from PCA, which demonstrates the superiority of HDFM.

\begin{table*}[tbp]
\centering
\setlength{\tabcolsep}{5mm}
\caption{Test accuracy comparison of different classification algorithms with two signal extraction methods on different datasets.}
\begin{tabular}{cclcccc}
\toprule
Datasets & \multicolumn{2}{c}{Methods} 
& Accuracy & Recall & Precision & F1-score \\ 
\midrule
\multirow{8}{*}{Meeting room} 
& \multirow{2}{*}{RF} 
& PCA  & 0.7736 & 0.7830 & 0.7837 & 0.7805 \\
& & HDFM & 0.8306 & 0.8375 & 0.8333 & 0.8350 \\ 
& \multirow{2}{*}{XGBoost} 
& PCA  & 0.8014 & 0.8043 & 0.8133 & 0.8033 \\ 
& & HDFM & 0.8583 & 0.8654 & 0.8631 & 0.8634 \\ 
& \multirow{2}{*}{SVM} 
& PCA  & 0.8236 & 0.8305 & 0.8283 & 0.8288 \\
& & HDFM & 0.8750 & 0.8782 & 0.8819 & 0.8797 \\
& \multirow{2}{*}{DS-STAN} 
& PCA  & 0.9083 & 0.9128 & 0.9171 & 0.9110 \\
& & HDFM & \textbf{0.9639} & 0.9651 & 0.9636 & 0.9643 \\

\multirow{8}{*}{Office}
& \multirow{2}{*}{RF} 
& PCA  & 0.6948 & 0.7036 & 0.6867 & 0.6857 \\
& & HDFM & 0.7631 & 0.7714 & 0.7652 & 0.7643 \\ 
& \multirow{2}{*}{XGBoost} 
& PCA  & 0.7219 & 0.7283 & 0.7120 & 0.7142 \\ 
& & HDFM & 0.8167 & 0.8218 & 0.8163 & 0.8181 \\ 
& \multirow{2}{*}{SVM} 
& PCA  & 0.5646 & 0.5690 & 0.5561 & 0.5300 \\
& & HDFM & 0.6512 & 0.6508 & 0.6418 & 0.6278 \\
& \multirow{2}{*}{DS-STAN} 
& PCA  & 0.9010 & 0.9011 & 0.8992 & 0.8991 \\
& & HDFM & \textbf{0.9726} & 0.9731 & 0.9738 & 0.9734 \\

\bottomrule
\end{tabular}
\label{table4}
\end{table*}

\subsection{Performance Evaluation of Amplitude-phase Fusion}
\begin{figure}[htbp]
  \centering
  \subfloat[Performance comparisons of different methods on the Meeting room dataset.]{
  \includegraphics[width=0.8\columnwidth]{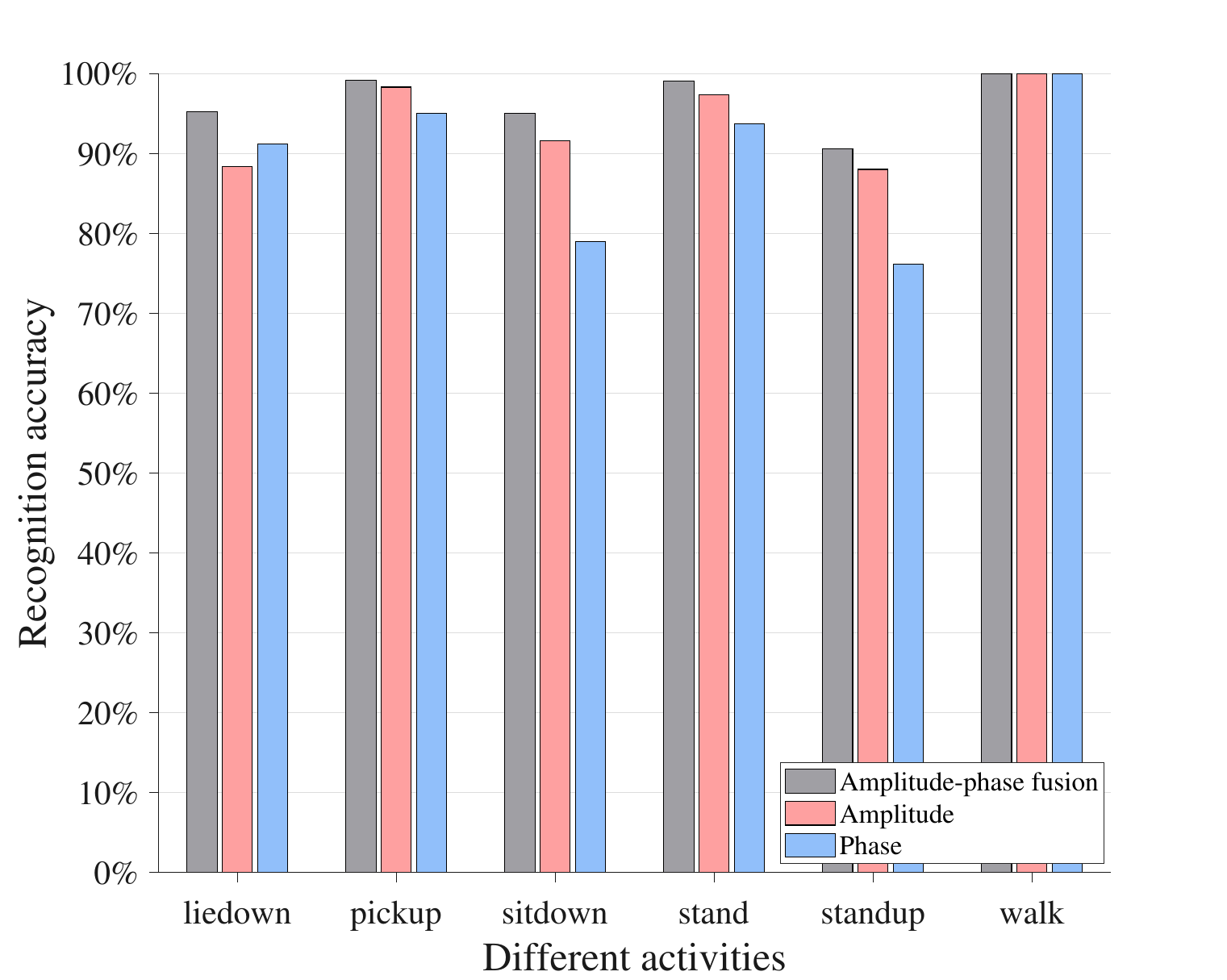}}
  \\
  \subfloat[Performance comparisons of different methods on the Office dataset.]{
  \includegraphics[width=0.8\columnwidth]{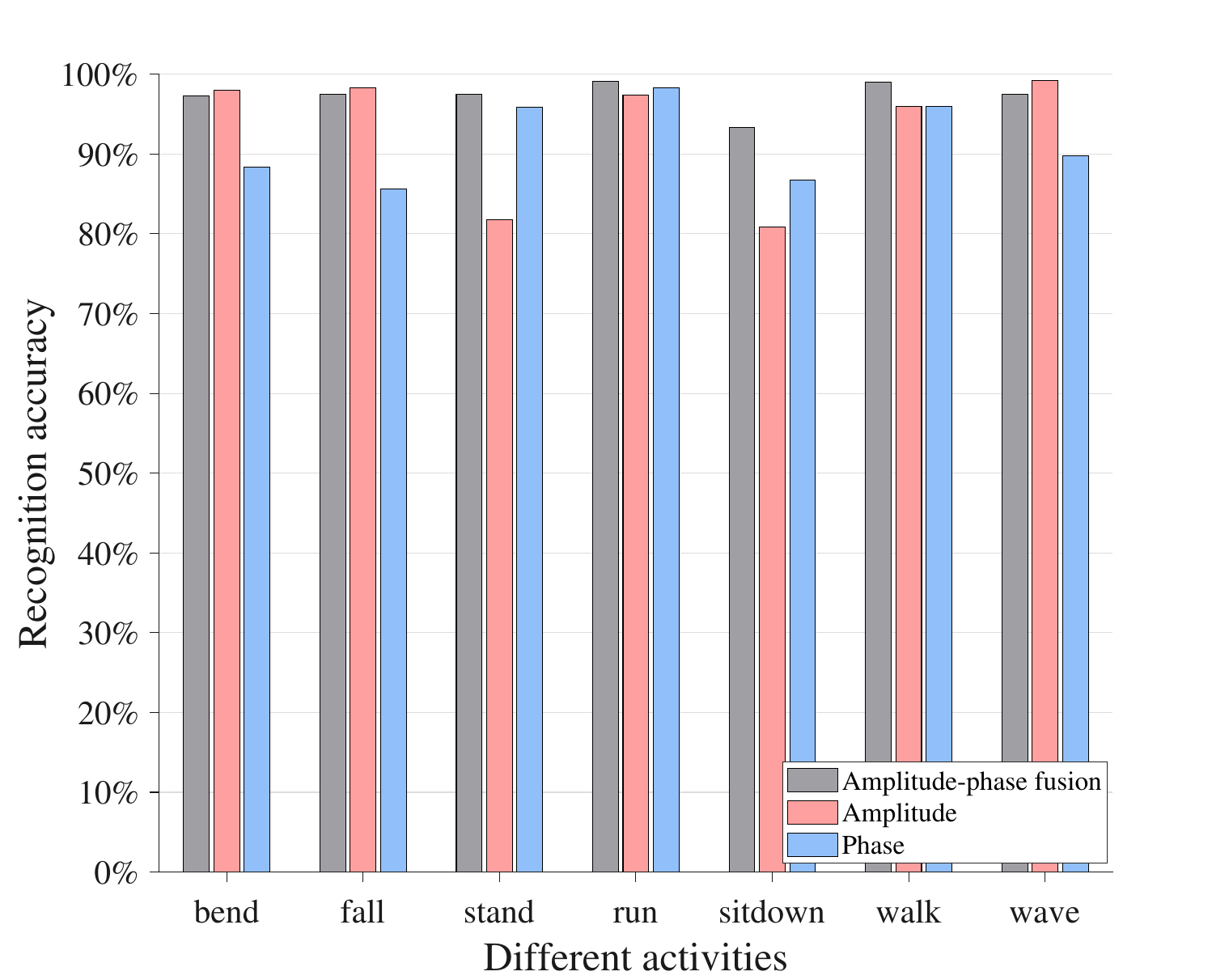}}
  \caption{Performance comparisons of different methods on the two different datasets.}
  \label{F6}
\end{figure}

The comparative analysis reveals a significant performance boost in human activity recognition through the application of CSI amplitude-phase fusion data, as depicted in Fig.~\ref{F6}. This fusion approach outperforms amplitude-only and phase-only methods, demonstrating higher accuracy across different datasets. Notably, the Meeting room and Office datasets show accuracy of 96.39\% and 97.26\% respectively with the fusion method, affirming the hypothesis that combining amplitude and phase data substantially enhances recognition accuracy. This academic consolidation underlines the pivotal role of amplitude-phase data fusion in advancing CSI-based activity recognition systems.

\subsection{Performance Evaluation of Attention Mechanism}
\begin{figure}[htbp]
\centering
\subfloat[Performance comparisons of with/without attention mechanism on the Meeting room dataset.]{
\includegraphics[width=0.8\columnwidth]{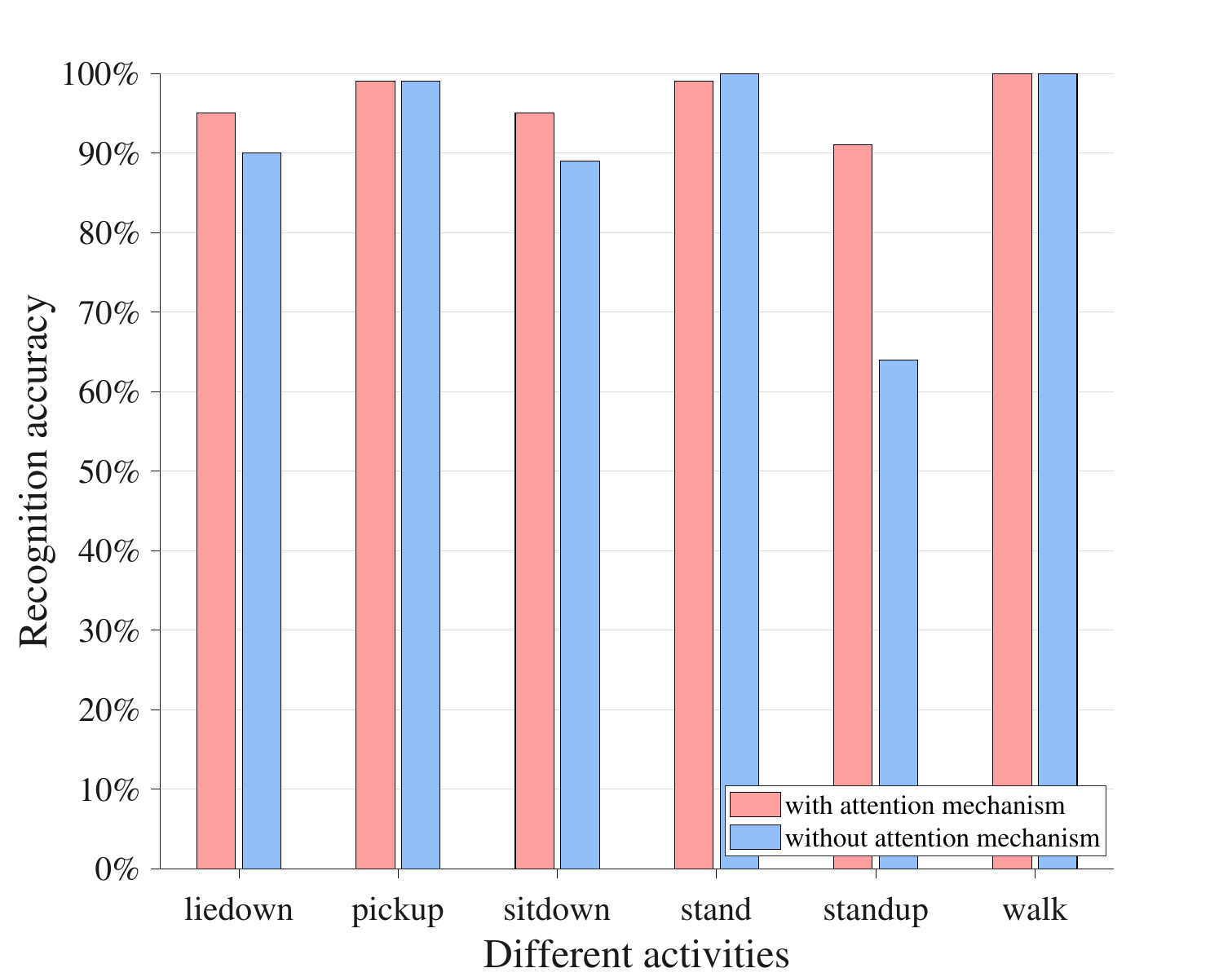}
}
\\
\subfloat[Performance comparisons of with/without attention mechanism on the Office dataset.]{
\includegraphics[width=0.8\columnwidth]{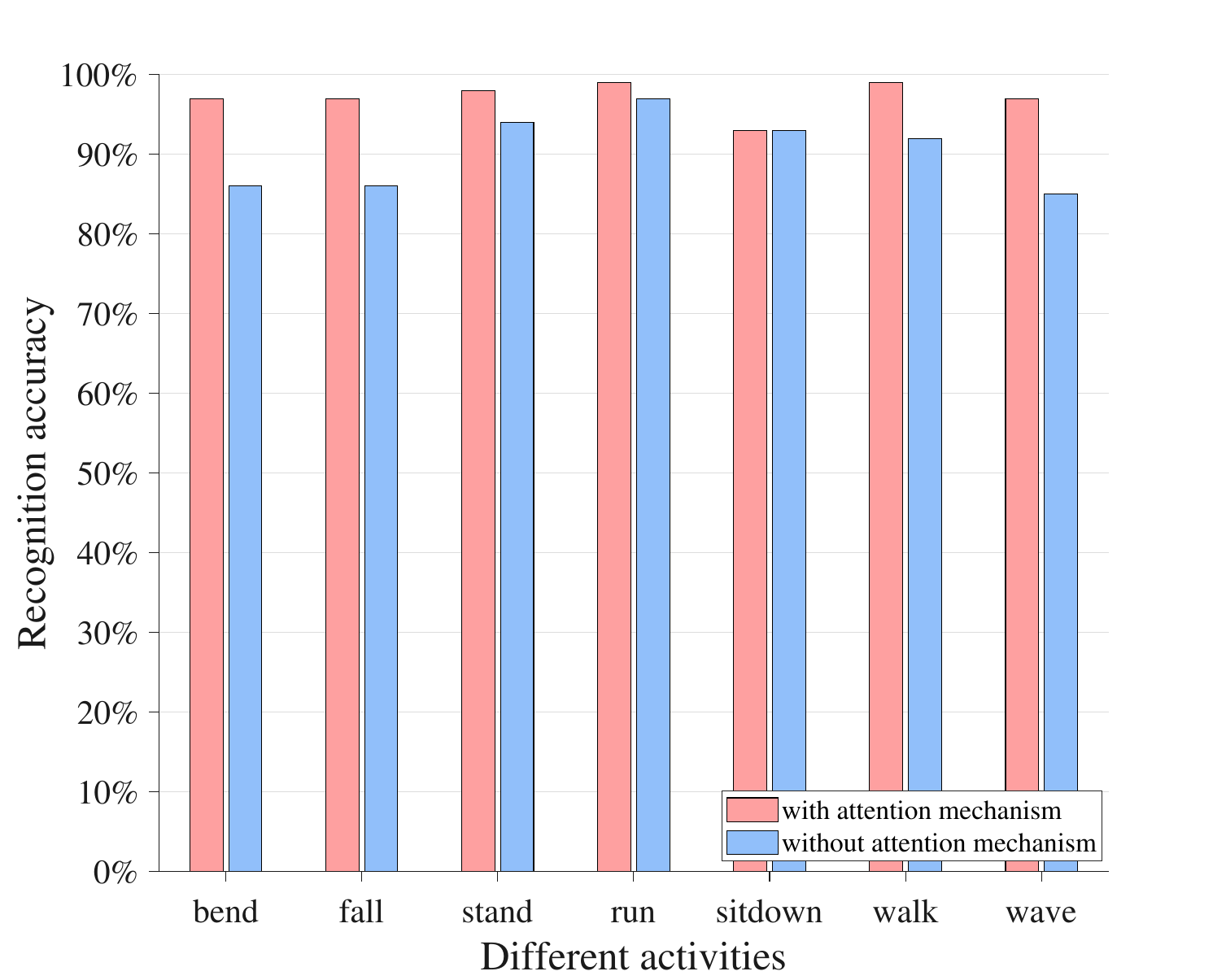}
}
\caption{Performance comparisons of attention mechanism on the two different datasets.}
\label{F7}
\end{figure}

Fig.~\ref{F7} reveals a pronounced improvement in activity recognition accuracy when utilizing a CSI amplitude-phase feature fusion method enhanced by an attention mechanism. Specifically, employing this mechanism increases accuracy to 96.39\% and 97.26\% for the Meeting room and Office datasets, respectively, a notable improvement from 90.14\% and 90.12\%. The attention mechanism's impact is particularly pronounced in the Meeting room dataset, where the accuracy for the stand-up activity soared from 64\% to 91\%. These results highlight the critical advantage of integrating an attention layer, showcasing not only an increase in accuracy and stability but also the method's adaptability and consistent performance across varied settings.

\subsection{Performance Evaluation with Other Existing Methods}
In our comparative analysis, presented in Table~\ref{table5}, the performance of our proposed DS-STAN method for activity recognition, utilizing the Meeting room dataset, significantly outperforms existing methods including LSTM, ABLSTM, Wi-multi, DeepSeg, LCED, CeHAR, and CNN-GRU-CNN, achieving an exceptional accuracy rate of 96.39\%. The result not only affirms the superior efficacy of our approach but also highlights its potential to emerge as a forefront solution in the HAR domain.

\begin{table}[htbp]
\centering
\caption{Performance comparison of the proposed approach with other existing methods.}
\setlength{\tabcolsep}{5mm}
\begin{tabular}{cccc}
\toprule
Authors & Methods & Accuracy \\
\midrule
Yousefi \textit{et al.} \cite{yousefi2017survey} & LSTM & 0.8708 \\

Chen \textit{et al.} \cite{chen2018wifi} & ABLSTM & 0.7139 \\

Feng \textit{et al.} \cite{feng2019wi} & Wi-multi & 0.7514 \\

Xiao \textit{et al.} \cite{xiao2020deepseg} & DeepSeg & 0.7389 \\

Guo \textit{et al.} \cite{guo2021towards} & LCED & 0.8681 \\

Lu \textit{et al.} \cite{lu2022cehar} & CeHAR & 0.9000 \\

Shalaby \textit{et al.} \cite{shalaby2022utilizing} & CNN-GRU-CNN & 0.9194 \\

Ours & RISAR & \textbf{0.9639} \\
\bottomrule
\end{tabular}
\label{table5}
\end{table}

\section{Conclusion}
This paper presents RISAR, a novel passive RIS-assisted human activity recognition system leveraging Wi-Fi signals, particularly optimized for low signal-to-noise ratio environments. By integrating a high-dimensional factor model for signal extraction and noise reduction with a dual-stream spatial-temporal attention network, RISAR significantly outperforms traditional PCA-based methods in accuracy and efficiency, achieving up to 97.26\% accuracy in experimental settings. This system's robustness, demonstrated across diverse environmental conditions, underlines its potential for wide-ranging IoT applications, from smart homes to healthcare. The study not only contributes substantially to the fields of wireless sensing and activity recognition but also paves the way for future exploration into combining RIS technology with advanced machine learning techniques, marking a significant step forward in realizing more intuitive and seamless smart environments.

\appendices
\section{Marčenko-Pastur law}
Let $\mathbf{X}=\{ x_{ij} \}$ be a $N \times T$ random matrix, whose entires are independent identically distributed (i.i.d.) variables with the mean $\mu(x)=0$ and the variance $\sigma^2(x) < \infty $. The corresponding covariance matrix is defined as $\mathbf{C} = \frac{1}{T} \mathbf{X}\mathbf{X}^H$. As $N, T \longrightarrow \infty $ but $c = \frac{N}{T} \in (0, 1]$, according to M-P law, the empirical spectrum distribution of $\mathbf{C}$ converges to the limit with probability density function (PDF)
\begin{equation}
f(x)=\begin{cases}
\frac{1}{2\pi c \sigma^2 x} \sqrt{(b-x)(x-a)}, & a \leq x \leq b \\
0, & \text{others}
\end{cases}
\end{equation}
where $a = \sigma^2(1-\sqrt{c})^2$, $b=\sigma^2(1+\sqrt{c})^2$.

\section{Proof of the HDFM}
\label{proof}
\newtheorem{assumption}{Assumption}
\begin{assumption}
\label{assumption1}
Let $\lambda_{1, T} \geq \ldots \geq \lambda_{p, T}$ represent the non-zero eigenvalues of $(\mathbf{L}\mathbf{F}) (\mathbf{L}\mathbf{F})^H$. For $j=1,\ldots,p$, $\lambda_{j, T} \rightarrow \lambda_j$ as $T \rightarrow \infty$, where
\[ \lambda_1 > \ldots > \lambda_p > \sigma^2 \sqrt{c}. \]
\end{assumption} 

\newtheorem{theorem}{Theorem}
\begin{theorem}
\label{theorem1}
Let $\hat{\lambda}_{1, T} \geq \ldots \geq \hat{\lambda}_{N, T}$ denote the eigenvalues of matrix $\mathbf{R} \mathbf{R}^H$. Under Assumption~\ref{assumption1}, for $j=1,\ldots,p$,
\[ \hat{\lambda}_{j, T} \xrightarrow[T \rightarrow \infty]{a. s.} \frac{(\lambda_j + \sigma^2)(\lambda_j + \sigma^2 c)}{\lambda_j}. \]
Moreover,
\[ \hat{\lambda}_{p+1, T}, \ldots, \hat{\lambda}_{N, T} \in [\sigma^2 (1-\sqrt{c})^2, \sigma^2(1+\sqrt{c})^2],  \]
almost surely (a.s.) for $T$ large enough.
\end{theorem}

Theorem~\ref{theorem1} is derived from the general results established in \cite{baik2006eigenvalues}. It delineates a sudden transition in the behavior of the $j$-th dominant eigenvalue $\hat{\lambda}_{j, T}$ of $\frac{1}{T}\mathbf{R} \mathbf{R}^H$: if $\lambda_{j, T} \leq \sigma^2 \sqrt{c}$, $\hat{\lambda}_{j, T}$ converges to the right-edge $\sigma^2(1+\sqrt{c})^2$ of the support defined by the Marčenko-Pastur law $\mu$, exhibiting no isolation. However, once $\lambda_{j, T} > \sigma^2 \sqrt{c}$, $\hat{\lambda}_{j, T}$ converges to a limit outside the right edge of $\mu$, indicating isolation from the Marčenko-Pastur support.

\bibliographystyle{IEEEtran}
\bibliography{Reference}

\newpage

\vfill

\end{document}